# The Chemical Effect Goes Resonant – A Full Quantum Mechanical Approach on TERS

Kevin Fiederling[a], Mostafa Abasifard,[a] Martin Richter,[a] Volker Deckert,[a,b,c], Stefanie Gräfe[a]*, Stephan Kupfer[a]*

[a] *Institute of Physical Chemistry and Abbe Center of Photonics, Friedrich Schiller University Jena, Helmholtzweg 4, 07743 Jena, Germany*

[b] *Leibniz Institute of Photonic Technology (IPHT), Albert-Einstein-Straße 9, 07745 Jena, Germany*

[c] *Institute of Quantum Science and Engineering, Texas A&M University, College Station, TX 77843-4242, USA*

Corresponding authors:

s.graefe@uni-jena.de

stephan.kupfer@uni-jena.de

# Table of content

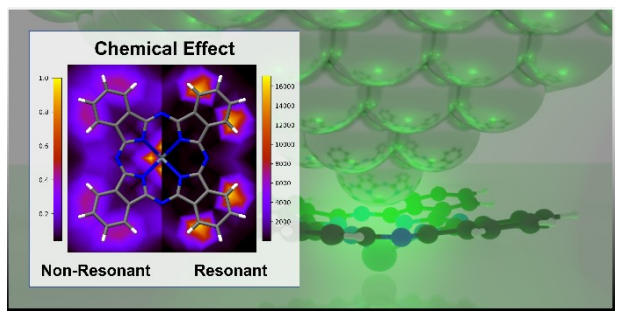

The lately postulated Angstrom resolution induced by non-resonant and resonant chemical interaction as well as by charge-transfer phenomena in plasmon-enhanced spectroscopies, *i.e.* in the scope of tip-enhanced Raman spectroscopy, was evaluated by a full quantum chemical approach.

# Keywords

TERS, quantum mechanics, chemical effect, non-resonant, resonant, lateral resolution

# Abstract

Lately, experimental evidence of unexpectedly extremely high spatial resolution of tip-enhanced Raman scattering (TERS) has been demonstrated. Theoretically, two different contributions are discussed: an electromagnetic effect, leading to a spatially confined near field due to plasmonic excitations; and the so-called chemical effect originating from the locally modified electronic structure of the molecule due to the close proximity of the plasmonic system. Most of the theoretical efforts have concentrated on the electromagnetic contribution or the chemical effect in case of non-resonant excitation. In this work, we present a fully quantum mechanical description including non-resonant and resonant chemical contributions as well as charge-transfer phenomena of these molecular-plasmonic hybrid system at the density functional and the time-dependent density functional level of theory. We consider a surface-immobilized tin(II) phthalocyanine molecule as the molecular system, which is minutely scanned by a plasmonic tip, modeled by a single silver atom. These different relative positions of the Ag atom to the molecule lead to pronounced alterations of the Raman spectra. These Raman spectra vary substantially, both in peak positions and several orders of magnitude in the intensity patterns under non-resonant and resonant conditions, and also, depending on, which electronic states are addressed. Our computational approach reveals that unique – non-resonant and resonant – chemical interactions among the tip and the molecule significantly alter the TERS spectra and are mainly responsible for the high, possibly sub-Angstrom spatial resolution.

# 1. Introduction

The holy grail driving the development of modern microscopic techniques is to reach the ultimate spatial resolution, allowing direct measurement of the sample molecule. This would necessitate Angstrom-scale spatial resolution – almost three orders of magnitude smaller than the Abbe diffraction limit for optical frequencies. Several modern techniques have been developed, ranging from fluorescence-based techniques as stimulated-emission-depletion (STED),[1] photoactivated localization microscopy (PALM)[2] and stochastic optical reconstruction microscopy (STORM),[3] to near-field techniques such as scanning near-field optical microscopy (SNOM).[4–8] The principle of the near-field optical methods relevant here, relies on the collectively oscillating electrons of a nanoscale metallic particle with the frequency of the incidental electromagnetic field, *i.e.* excitation of a (surface-)plasmon. This oscillating charge gives rise to a locally confined and enhanced electromagnetic field.

Among the most prominent plasmon-based techniques are surface- and tip-enhanced Raman scattering (SERS and TERS, respectively).[9–12] Such techniques commonly allow an increasement of the Raman signal by up to six orders of magnitude,[13,14] which can be rationalized by two contributions: the electromagnetic effect[15–23] and the chemical effect.[16,24–29] These effects can be specified as follows: the electromagnetic effect leads to a pronounced increase in the corresponding Raman signals of a sample near the plasmonic particle by the locally confined and enhanced electric field. It can be substantially enhanced when making use of pico-cavities created when anchoring the molecule between the plasmonic tip and the (conducting) substrate.[30–34] Additionally, the interaction between the dipole induced in the molecule by the enhanced field and the corresponding mirror dipole in the plasmonic nanoparticle leads to a modified nearfield, and, consequently, to a substantially narrowed nearfield-profile.[35–37] The chemical effect, on the other hand, originates from close-range interactions between the molecular sample and the metallic nanoparticle. By scanning with the tip over the molecule, different tip-sample interactions occur, leading to variations of the spectral position and the intensity of Raman bands due to the minute changes in the local chemical environment of the sample.[17,38] In general, the chemical effect originates from: i) non-resonant contributions, *i.e.*, ground-state interactions between the sample and the nanoparticle, ii) resonant contributions emerging from excited-states of the sample (molecule) in the vicinity of the excitation wavelength of the electric field, and iii) from charge-transfer phenomena between the sample and the nanoparticle (charge-transfer contribution).[16,24,39–42]

Recently, strong evidence was obtained by TERS experiments – both under ambient and cryogenic conditions – indicating that lateral resolutions near or even below 1 nm are possible,[43–50] thus enabling single molecule – or even sub-molecular – resolution. Such resolutions are highly surprising, as radii of commonly applied plasmonic nanoparticles range between 10 and 20 nm.[51–54] Recent works suggest that the presence of single, atomic-sized features at the tip-apex (such as corners or edges) may provide an explanation for the surprisingly high resolution [31,33,55,56] and that only minor contributions can be attributed to deeper lying atomic layers.[38,57] However, the specific origin of the high spatial resolution is still under debate.

In this contribution, we analyze theoretically the different possible chemical effects listed above, investigating for the first time, how the Raman signal changes when going from non-resonant interaction to resonant interactions of the molecular-plasmonic hybrid system. Common approaches for theoretical and numerical modeling of such plasmon-molecule hybrid systems were developed in the groups of Schatz[58–61], Jensen[33,54,62,62–65] and Aizpurua[12,34,43,55,56] and have advanced our current understanding. The most common model is the so-called discrete interaction model/quantum mechanical (DIM/QM) model developed by Jensen et al.[35,36,54,63] combining time-dependent density functional theory (TDDFT) simulations for the molecular sample with atomistic electrodynamics simulations for the nanoparticle. Furthermore, Aizpurua et al. and Sánchez-Portal et al. presented full quantum electrodynamic descriptions of plasmonic cavities, demonstrating the resolution capabilities of plasmonic nanoparticles due to extremely confined enhanced fields that can explain sub-nanometer resolution via the atom-scale features of small metal clusters.[55,56] Another approach by Luo et al. presents a quantum mechanical description of the interaction between the molecule and the metal tip, treated as a highly confined plasmonic field, using an electric field function entering an effective Hamiltonian.[66–68]

In a completely different approach, we have recently introduced a computational protocol describing the site-specific non-resonant molecular-plasmonic interactions entirely in a quantum chemical fashion.[38] This concept was applied to mimic the TERS signal of surface-immobilized adenine. It allowed us to elucidate the contribution of the non-resonant chemical effect, *i.e.* the local chemical environment of the molecule near the silver particle. This has been achieved by three-dimensional grid calculations performed at the density functional (DFT) level of theory mapping the adenine molecule with a single silver atom or a small silver cluster, mimicking a fine scanning of the plasmonic silver tip in all three spatial degrees of freedom over the molecule with minute precision. We could show that this mere non-resonant chemical interaction between the plasmonic tip and the adenine molecule results in pronounced local modifications of the Raman signal, both including spectral positions and intensities of the vibrational modes. Our calculations provided a possible explanation for the recently proposed molecular or even sub-molecular resolution of TERS under non-resonant conditions[49] to a certain extent.

In the current contribution, we present a full quantum chemical investigation addressing the highly site-specific nature of the chemical effect in a TERS setup, including non-resonant, resonant and charge-transfer contributions. We show that depending on the excitation wavelength, different electronic states are addressed, resulting in very different TERS spectra.

As a realistic model system, we chose the surface-immobilized molecule tin(II) phthalocyanine (**SnPc**) which is three-dimensionally mapped by a single silver atom employing DFT and TDDFT calculations. This class of molecules, (metallo-)phthalocyanines[69–72] and (metallo-)porphyrins,[43,48,64,73,74] has been subject to numerous experimental and theoretical studies in the scope of plasmon-enhanced spectroscopy (SERS and TERS). In particular, **SnPc** was selected for the present computational investigation as pronounced resonance effects are anticipated for **SnPc** stemming from bright absorption features in the visible region. It is the aim of this paper, to investigate resonance Raman (RR) contributions, *i.e.* via **SnPc**-centered ($\pi\pi^*$) states and charge-transfer states. Furthermore, **SnPc** can be (potentially) easily mapped by the silver particle and a tip-induced reorientation on the surface is highly conceivable due to its slightly conical structure. We calculate resonance effects depending on the local chemical environment, the wavelength of the incident electric field as well as on a potential tip-induced reorientation of the surface-immobilized **SnPc**.

With this approach, we can address for the first time the resonance Raman spectra as in a typical TERS setup. By tuning the excitation wavelength, the contribution of excited-states of the hybrid system in resonance with the exciting laser are addressed. We demonstrate that the simulated Raman intensity maps give insight regarding the local Raman response as well as aim on addressing the lateral resolution under non-resonant and resonant conditions.

## 2. Computational details

The computational protocol employed in the present study is based on our recently introduced setup,[38] where local "forced" molecule-tip interactions are simulated at the DFT level of theory by mapping an (surface-)immobilized molecule along a three-dimensional grid with a plasmonic nanoparticle mimic, *i.e.*, a small silver cluster or even a single silver atom.

All calculations were performed using the Gaussian 09 program.[75] The isolated tin(II) phthalocyanine (**SnPc**) of $C_{4v}$ symmetry was preoptimized at the density functional level of theory using the range-separated XC functional CAM-B3LYP[76] and the 6-311+G(d,p) triple-$\zeta$ basis set.[77,78] The tin atom was described using the relativistic electronic core potential MWB46[79] for the inner electrons, while the valence electrons are treated explicitly with double-$\zeta$ quality. All calculations were performed including D3 dispersion correction with Becke-Johnson damping.[80] A vibrational analysis confirmed that a minimum on the $3N-6$-dimensional potential energy (hyper-)surface (PES) was obtained. To correct for the lack of anharmonicity and for the approximate description of electron correlation, the harmonic frequencies were scaled by a factor of 0.95.[81]

Subsequently, **SnPc** is mapped by three-dimensional grid calculations, where the plasmonic tip was approximated by a single silver atom which is described using the relativistic MWB28[79] pseudo potential for inner electrons and the respective basis set (double-$\zeta$) for valence electrons. As shown by our group, one single silver atom is sufficient to mimic characteristic (non-resonant) chemical interactions in such hybrid systems.[38] Therefore, the aromatic system of the slightly conical molecule was aligned to the $xy$-plane, with the tin atom facing downwards at the center of the coordinate system at $z = 0$, see Figure 1A.

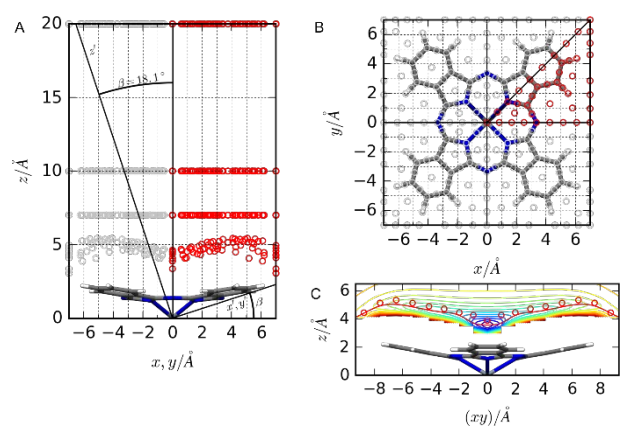

**Figure 1:** A: View along the $x$- or $y$-axis showing the $z$-values of the different scan layers and the possible tip-induced reorientation ($\angle_{\tau}$ = 18.1°). Circles indicate $x$, $y$-scanning positions of the Ag-atom; red circles indicate the calculated 41 positions, gray circles represent projected points using the molecular symmetry ($C_{4v}$). B: View along the z-axis with **SnPc** in the *xy*-plane. C: Potential energy landscape obtained by partial structural relaxation, *i.e.* the *z*-coordinate of **SnPc-Ag** in the 17 *xy*-positions along the diagonal; one contour level corresponds to an energy difference of 0.05 eV. Tip positions based on the van-der-Waals (vdW) radii (red circles) are compared to the relaxed **SnPc-Ag** bond distances (red line).

By taking advantage of **SnPc**'s molecular symmetry ($C_{4v}$), mapping a slice of one-eighth is sufficient to recover the entire map by applying a sequence of symmetry operations. Thus, 41 preselected points along the *xy*-coordinate space, such as atomic positions and centers of bonds and rings, were investigated. The tip-sample distance is given by the *z*-value, therefore, several equidistant layers at *z* = 20, 10, and 7 Å as well as a layer based on the sum of the van-der-Waals (vdW) radii of **SnPc** and the scanning silver atom were constructed, see Figure 1A and B for a detailed description of the applied grid. Due to the odd number of electrons of the silver atom ($4d^{10}\ 5s^1$), the electronic ground-state of **SnPc-Ag** is a doublet, where the 5s orbital of the silver is singly occupied.

We mimic the TERS setup as follows: The surface-immobilized sample (**SnPc**) is aligned in the *xy*-plane with the tip being vertically above. As the tip is in *z*-direction, this is the direction along which Raman enhancement occurs dominantly. Consequently, only the $\alpha_{zz}$-component of the polarizability tensor is subjected to the electromagnetic enhancement (assuming parallel field vectors). Within the so-called sum-over-states expression as derived by Kramers, Heisenberg and Dirac $\alpha_{zz}$ is given by:[82,83]

$$(\alpha_{zz})_{i \to f} = \sum_n \left( \frac{\langle f|\hat{\mu}_z|n\rangle \langle n|\hat{\mu}_z|i\rangle}{E_{n,i} - E_L - i\Gamma} + \frac{\langle f|\hat{\mu}_z|n\rangle \langle n|\hat{\mu}_z|i\rangle}{E_{n,f} + E_L + i\Gamma} \right) \quad , \tag{1}$$

where $\hat{\mu}_z$ is the *z*-component of the dipole moment operator, $E_{n,i}$ the Bohr energy for a transition between the two vibronic states *i* and *n* defined as $E_{n,i} \equiv E_n - E_i$, $E_L$ the energy of the irradiating laser and $\Gamma$ the damping factor describing homogeneous broadening (0.372 eV). Eq. (1) comprises a resonant term (first) and a non-resonant term (second). The latter term as well as its derivative $\partial(\alpha_{zz})_{i \to f}/\partial q_l$ is obtained by (ground-state) frequency calculations within the harmonic approximation for a fundamental transition ($g0_l \to g1_l$), *i.e.*, a transition from the vibrational ground-state of the *l*th vibrational mode of the electronic ground-state ($g0_l$) to its first vibrational excited state ($g1_l$). Under non-resonant conditions, the Raman intensity of the *l*th vibrational normal mode $I_l$ is approximately proportional to the forth power of the energy difference of the irradiating laser ($E_L$) and the respective normal mode ($E_l$) times the absolute square of the partial derivative of $\alpha_{zz}$ with respect to the normal coordinate ($q_l$):

$$I_l \propto (E_L - E_l)^4 \left| \frac{\partial (\alpha_{zz})_{g0_l \to g1_l}}{\partial q_l} \right|^2 . \qquad (2)$$

Hence, the integrated, or rather summed, signal intensity *I* at each grid-point (*x*, *y*, *z*) is given by:

$$I \propto \sum_l I_l = \sum_l (E_L - E_l)^4 \left| \frac{\partial (\alpha_{zz})_{g0_l \to g1_l}}{\partial q_l} \right|^2 . \qquad (3)$$

The resonance term of eq. (1) is obtained by assuming exclusively Condon-type scattering (A-term contribution[84]) for a fundamental transition within the transform theory:[85–90]

$$(\alpha_{zz})_{g0_l \to g1_l} = \sum_e (\mu_z)_{ge}^2 \frac{\Delta_{e,l}}{\sqrt{2}} \left( \Phi_e(E_L) - \Phi_e(E_L - E_l) \right) , \qquad (4)$$

where the function $\Phi_e$ is given by:[88,91,86]

$$\Phi_e(E_L) = \frac{1}{E_{g,e} - E_L - i\Gamma} . \qquad (5)$$

Here $E_{g,e}$ is the vertical excitation energy from the electronic ground-state *g* to the excited-state *e*; Franck-Condon factors were neglected. Furthermore, the dimensionless displacement $\Delta_{e,l}$ of the excited-state *e*, in eq. (4), is defined in the independent-mode displaced harmonic oscillator model, which assumes the electronic ground- and excited-state potentials to be spatially displaced harmonic oscillators, sharing the same set of vibrational modes, given by the partial derivative of the excited-state potential energy $E_e$ along the normal mode $q_l$ in the electronic ground-state equilibrium geometry:

$$\Delta_{e,l} = \frac{-\hbar^2}{\sqrt{E_l^3}} \left( \frac{\partial E_e}{\partial q_l} \right)_0 \quad . \tag{6}$$

In the following, the subscript $g0_l \rightarrow g1_l$, indicating a fundamental transition, will be neglected for the sake of simplicity.

TDDFT simulations were performed employing the same XC functional as in the preliminary ground-state calculations, with the basis set being reduced from 6-311+G(d,p) to 6-31G(d)[92] for computational reasons in order to evaluate the properties of the bright excited-states contributing to the resonance term in eq. 1 for **SnPc-Ag**. While a highly flexible basis set is essential to accurately describe ground-state polarizabilities, transition polarizabilities – as obtained by the present methodology – merely depend on vertical excitation energies, excited-state gradients and transition dipole moments, hence, a computationally less demanding double-ζ basis is sufficient.[86,93–97] We confirmed by direct comparison of both basis sets that the excited-state properties of **SnPc** are hardly affected by the reduction of the basis set from triple-ζ to double-ζ (Table S1). In the following, the 20 lowest excited doublet states of **SnPc-Ag** were calculated at each point of the three-dimensional grid to assess excitation energies, transition dipole moments and analytical Cartesian energy derivates of the excited-states, exclusively using the computationally less demanding double-ζ basis set. To further reduce the computational demand, only excited-state gradients of bright excited-states (with a $z$-component of the oscillator strength of $f_z \geq 0.01\, a.u.$) in resonance at the applied excitation energy $E_L$ ($0 \leq E_{g,e} \leq 3.2\, eV$) were calculated.

Under non-resonant conditions ($E_{g,e} \neq E_L$), the $z$-polarized Raman signal was obtained by ground-state frequency calculations yielding the polarizability derivatives, while an excitation energy of $E_L$ = 1.166 eV ($\lambda_L$ = 1064 nm) was assumed to yield $I_l$ as in eq. 2. The $z$-polarized Raman intensities under resonant conditions ($E_{g,e} \approx E_L$) at $E_L$ = 1.959, 2.331 and 2.807 eV ($\lambda_L$ = 633, 532 and 442 nm) are given by the respective transition polarizability derivatives, while the much weaker non-resonant contributions were neglected for these resonant conditions (recall eq. 1). All $z$-polarized Raman signals were broadened by Lorentzians with a full width at half maximum (FWHM) of 5 cm$^{-1}$.

Finally, $z$-polarized non-resonant and resonant Raman signals were obtained to assess the impact of a tip-induced "mechanical" reorientation. To mimic this, **SnPc-Ag** was tilted by an angle of ß = 18.1° along the $y$-axis, as indicated in Figure 1A. The molecular properties were then obtained by a unitary transformation:

$$\alpha' = R_y \alpha R_y^T \qquad , \qquad (7)$$

with

$$R_y = \begin{pmatrix} \cos\beta & 0 & \sin\beta \\ 0 & 1 & 0 \\ -\sin\beta & 0 & \cos\beta \end{pmatrix} \qquad , \qquad (8)$$

where $R$ is a rotational matrix and $\alpha'$ the new rotated polarizability tensor. Accordingly, $\alpha'_{zz}$ – the $zz$-component of the rotated polarizability tensor – was used to calculate the non-resonant and resonant polarizability derivatives and Raman intensities with respect to eq. 2 for the reoriented hybrid system.

# 3. Results and discussion

The following section presents the simulated TERS spectra of the plasmonic hybrid system in a full quantum mechanical fashion, while the chemical effect is investigated – comprising non-resonant, resonant effects and charge-transfer phenomena at the tip-substrate interface. Furthermore, the lateral resolution, *i.e.*, the possibility of molecular or even sub-molecular resolution, is thoroughly elaborated by three-dimensional grid calculations performed at the DFT and TDDFT levels of theory.

## 3.1. Non-resonant contribution

Initially, the TERS spectra of **SnPc-Ag** are calculated for the case of a non-resonant excitation. Therefore, the plasmon-enhanced Raman response of the molecule was obtained at each point along the three-dimensional grid introduced in section 2 (Figure 1). For the present study several equidistant $z$-layers, at 20, 10 and 7 Å as well as the vdW-layer were investigated, see Figure 1A. In the following, we focus our computational analysis mainly on the equidistant layer at $z = 7$ Å as well as on the vdW-layer, while the 20 Å-layer is used as reference for the non-interacting hybrid system (almost $z \to \infty$); all mode and integrated signal intensity enhancements described in the following are obtained with respect to this non-interactive reference layer. The $z$-dependency of the non-resonant signal was already studied previously in-depth.[38]

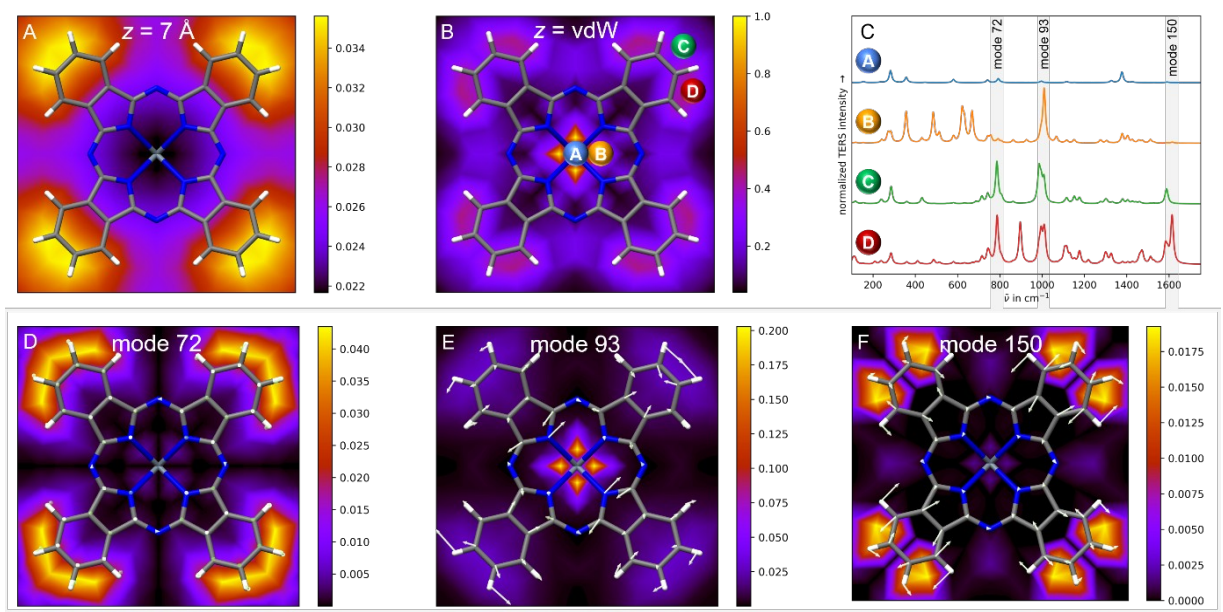

**Figure 2:** A and B: Integrated intensity maps of the *z*-polarized Raman spectra of the hybrid system. C: *z*-polarized Raman signals at four selected tip positions (**A**, **B**, **C** and **D**), as indicated in B. Intensity maps of prominent vibrational modes contributing to the Raman spectra are illustrated in D: out-of-plane CH bending, E: ring breathing, F: CC stretching, displacement vectors are shown in white.

While **SnPc** mapped at $z = 20$ Å yields nearly identical *z*-polarized Raman signals and integrated intensities ($I = 0.0244\,arb.u. \pm 0.0002\,arb.u.$, recall eq. 3), lowering the silver tip to 7 Å induces site-specific **SnPc** silver interactions. As expected, the integrated intensity map depicted in Figure 2A reveals that the highest intensities of up to $0.0356\,arb.u.$ are predicted in the vicinity of the CH groups, which equals roughly to an signal enhancement of 1.5. This finding is in agreement with the previous investigation on the non-resonant contribution of the adenine molecule.[38] However, in contrast to the planar adenine, the structure of **SnPc** is slightly conical, therefore, the plasmonic tip interacts at $z = 7$ Å more effectively with the upward pointing CH groups than with the central aromatic backbone or even the downward pointing central Sn atom. Furthermore, not only CH out-of-plain modes contribute to the *z*- polarized Raman spectra but also in-plain CH bending modes, due to their (partial) *z*-polarization.

From 7 Å the tip approaches further to the molecule until the minimum of the PES is reached within each *xy*-position of the map. This surface – representing the region of maximum attractive (chemical) interactions between **SnPc** and the silver tip – was approximated by the vdW-radii of the respective atoms. In order to estimate the accuracy of this approximation, the tip-sample distance, *i.e.*, the *z*-distance between the silver atom and **SnPc**, was optimized within all 41 positions along the map. The comparison of the PESs, based on the vdW-radii and the (partial) relaxations, yields a good agreement of both models, while a slight overestimation of the **SnPc-Ag** bonding distance of up to ≈0.3 Å is obtained for the computationally less demanding vdW-layer, see Figure 1C. In the following, all TERS simulations are performed at the vdW-layer.

At the vdW-layer, a further increase of the Raman intensity is calculated in the region of the CH bonds from 0.0356 (at 7 Å, Figure 2A) up to $1\,arb.u.$ or rather an enhancement of 41.0 (vdW, Figure 2B) – in-line with the previously predicted exponential signal amplification with the tip aproaching the sample.[38] In the vicinity of the CH bonds, as shown exemplarily in Figure 2C for the *z*-polarized Raman spectra in the positions **C** and **D**, mainly in-plain and out-of-plain CH modes, *e.g.,* modes 72 and 150 at ≈790 and ≈1615 cm$^{-1}$, contribute to the integrated intensity map, see Figures 2E and F. However, the highest integrated Raman signals are predicted in position **B** – adjacent to the central Sn atom. These signals mainly stem from a ring breathing mode (mode 93 at 1010 cm$^{-1}$), as evident from the Raman spectrum at position **B** (Figure 2C) as well as by means of the intensity map of the respective mode, shown in Figure 2E. In position **A**, with the tip directly above the tin atom, and in close proximity to the bright position **B**, the signal decreases rapidly to merely $0.0468\,arb.u.$, Figure 2C. The reason for this low intensity originates from the fact that many partial derivatives of $\alpha_{zz}$ along the respective normal modes are zero ($\partial \alpha_{zz}/\partial q_l = 0$) due to the high symmetry of **SnPc-Ag** ($C_{4v}$) at this grid point, Figure 2C. Consequently, the positions **A** and **B**, which are less than one Angstrom apart, can be resolved by using merely the integrated intensity map – without the need to compare the individual Raman spectra. The comparison of these Raman spectra and their underlying vibrational modes within each grid point of the integrated intensity map, exemplarily shown in Figure 2C for the positions **A**, **B**, **C** and **D**, reveals that the vibrational frequencies of the modes are nearly independent of the tip-position. However, the intensity as well as the intensity pattern is very sensitive to the tip position, reflecting the individual chemical environment of the molecular-plasmonic hybrid system. Therefore, our simulations provide a certain explanation for the (sub-)molecular resolution of TERS as predicted by recent experimental studies under non-resonant conditions.[48,49]

Based on the performed three-dimensional grid calculations, it is further possible to determine the signal amplification at each grid point and for each vibrational mode upon approach of the tip (with respect to the non-interacting reference layer at $z = 20$ Å). For the integrated intensity at 7 Å, slight signal amplifications by a factor of up to 1.4 are predicted in the region of the CH bonds, while a further increase of the amplification of up to 40 is calculated at the vdW-layer. The latter pronounced amplification is observed in the proximity of the central Sn atom. However, for certain modes amplifications of several orders of magnitude are obtained, *e.g.* at the vdW-layer the stretching of the SnN$_4$ fragment (mode 93) and the CH bending (mode 150) are increased by a factor of $1.3 \times 10^9$ and more than $1.6 \times 10^{10}$, respectively. Further details regarding the vibrational modes contributing to the non-resonant signal at the vdW-layer are collected in Figure S1 of the supporting information.

**3.2 Resonant and charge-transfer contributions**

In this section, we focus now on the resonance effects, *e.g.* for laser frequencies resonant to one or several (bright) electronic transitions of the hybrid system. To evaluate the influence of resonance enhancement, *i.e.* by $\pi\pi^*$ states of **SnPc** and charge-transfer states between the sample and the plasmonic tip, on the (*z*-polarized) Raman spectra, intensity maps, amplifications and ultimately the impact on the lateral resolution, in a full quantum chemical picture, it is essential to investigate the excited-state properties along the three-dimensional grid. Therefore, TDDFT simulations were carried out within each grid point to calculate the transition polarizabilities according to eq. 4 – based on excitation energies ($E_{g,e}$), transition dipole moments ($\mu_{ge}$) and Cartesian energy derivatives of the excited-states ($\left(\partial E_e / \partial q_l\right)_0$).

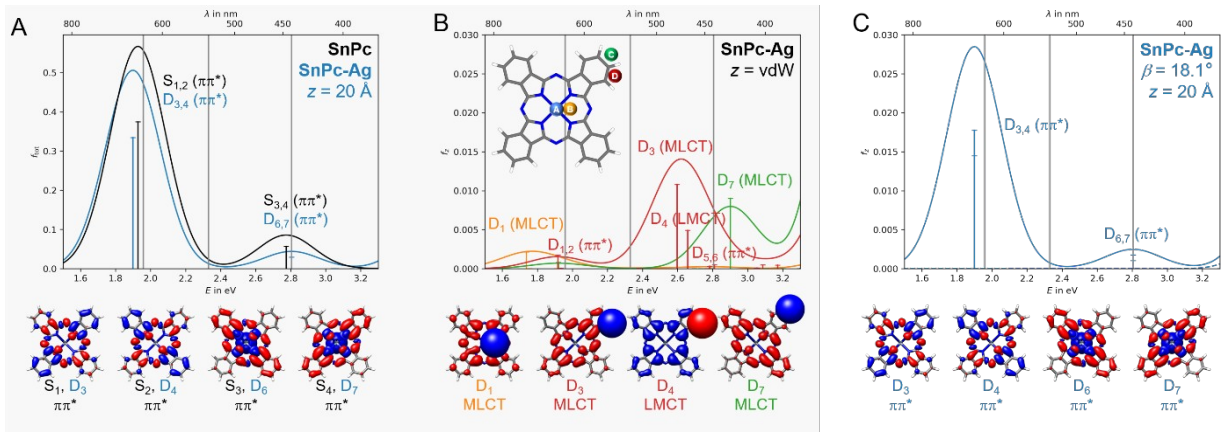

**Figure 3:** A: UV/vis absorption spectra of **SnPc** (in black) and **SnPc**-**Ag** at *z* = 20 Å (in blue, position **A**) and charge density differences (CDDs) for contributing $\pi\pi^*$ states (**SnPc**: $S_1$-$S_4$, **SnPc**-**Ag**: $D_3$-$D_7$); charge-transfer takes place from blue to red. B: *z*-polarized UV/vis spectra of the surface-immobilized **SnPc** mapped by one silver atom at the van-der-Waals (vdW) layer at selected tip positions. CDDs are illustrated for charge-transfer states. C: *z*-polarized UV/vis absorption spectrum of the tilted **SnPc**-**Ag** (in solid blue, position **A**, $\beta$ = 18.1°) and of the untilted **SnPc**-**Ag** (in dashed blue, position **A**, $\beta$ = 0°) hybrid systems, both at *z* = 20 Å. CDDs for $\pi\pi^*$ states, optically accessible upon tip-induced reorientation, are illustrated. Gray vertical lines indicate the wavelengths of the incidental electric field ($\lambda_L$ = 633, 532 and 442 nm; $E_L$ = 1.959, 2.331 and 2.807 eV).

The UV/vis absorption spectrum of **SnPc** in the absence of silver, see Figure 3A, is governed in the visible region by two pairs of degenerated bright $\pi\pi^*$ states of the phthalocyanine: $S_1$ and $S_2$ at 1.93 eV (643 nm) and $S_3$ and $S_4$ at 2.77 eV (447 nm), as indicated by the charge density differences (CDDs; Table S1). Upon inclusion of the silver tip model (silver atom: doublet with $4d^{10}\,5s^1$), a doublet ground-state is obtained for **SnPc-Ag**, with the UV/vis spectrum being hardly altered at $z = 20$ Å. A closer look at the $\pi\pi^*$ states ($S_{1-4}$ and $D_{3,4,6,7}$ at $z = 20$ Å) reveals that these states are polarized in *xy*-direction and do not feature a *z*-component. Consequently, no resonant enhancement is predicted at the non-interactive reference layer. However, reducing the tip-sample distance to the region of chemical interaction, *i.e.* to the vdW-layer, the four $\pi\pi^*$ excitations gain *z*-polarization. In consequence, the $\pi\pi^*$ states of the surface-immobilized **SnPc** are accessible at small tip-sample distances (Figure 3B). Pronounced local and state specific variations of the *z*-component of the transition dipole moment or rather its square $(\mu_z)_{ge}^2$ are induced by the plasmonic tip, *i.e.* three $\pi\pi^*$ states exhibit the highest $(\mu_z)_{ge}^2$-values at the periphery of **SnPc** (up to ≈0.040 a.u.), while the fourth locally excited **SnPc**-centered excitation is mainly accessible in proximity of the $SnN_4$ fragment ($(\mu_z)_{ge}^2 \leq 0.014\,a.u.$).

In addition to these locally excited-states, charge-transfer states among the silver atom and **SnPc** gain importance at small tip-sample distances. At the vdW-layer mainly two metal-to-ligand charge-transfer (MLCT) states, *i.e.* field-induced electron transfer from the tip to the sample, and one ligand-to-metal charge-transfer (LMCT) state, *i.e.* electron transfer from the sample to the plasmonic nanoparticle, contribute significantly to the site-specific UV/vis spectra of the hybrid system. In general, all charge-transfer states feature a pronounced absorption in *z*-direction, as charge migration predominantly occurs along the *z*-coordinate, see Figure 3B. The excitation energies of these charge-transfer states highly depend on the local chemical environment and varies by up to 2 eV along the vdW-layer; see Table S2 for details. Besides the energetic position of these charge-transfer states also their optical accessibility is found to be highly site-specific: The low-lying MLCT and LMCT states feature the highest $(\mu_z)_{ge}^2$-values in position **D**, while the higher lying MLCT states exhibit the strongest transition dipole moments in *z*-direction in position **C**, see Figure 3B and Table S2.

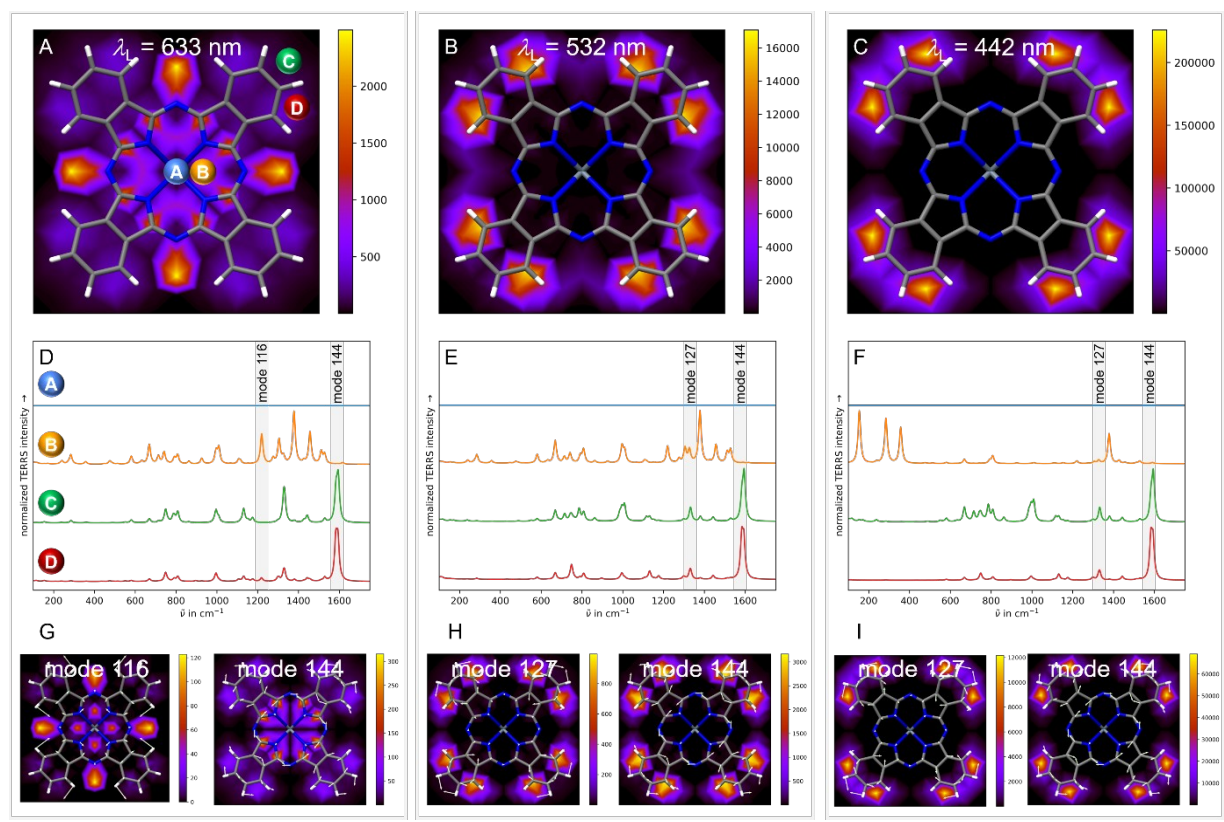

**Figure 4:** A-C: Integrated intensity maps of the *z*-polarized resonant Raman spectra of the hybrid system at the van-der-Waals (vdW) layer upon excitation at $E_L$ = 1.959, 2.331 and 2.807 eV ($\lambda_L$ = 633, 532 and 442 nm), respectively; four selected silver tip positions (**A**, **B**, **C** and **D**), are indicated. D-F: *z*-polarized resonance Raman signals at the positions **A**, **B**, **C** and **D**. G-I: Intensity maps of prominent vibrational modes contributing to the Raman spectra illustrated in D-F.

Based on the excited-state properties obtained by TDDFT, the $zz$-component of the transition polarizability tensor ($\alpha_{zz}$) was obtained assuming exclusively Condon-type scattering (eq. 4). Subsequently, the RR intensity of the $l$th vibrational normal mode $I_l$ is obtained under resonance of the $e$th excited-state with the incidental laser ($E_L$ = 1.959, 2.331 and 2.807 eV; $\lambda_L$ = 633, 532 and 442 nm) as in eq. 2. Under resonant conditions, this resonant term exceeds the non-resonant term by several orders of magnitude (eq. 1), therefore, the latter non-resonant term was neglected.

As discussed, the site-specific UV/vis spectra at the vdW-layer feature contributions from bright $z$-polarized charge-transfer states, *i.e.* of MLCT and LMCT character, as well as from locally excited $\pi\pi^*$ states of the surface-immobilized **SnPc**. At an excitation energy of $E_L$ = 1.959 eV ($\lambda_L$ = 633), the integrated $z$-polarized RR intensity map (Figure 4A) comprises contributions of MLCT and $\pi\pi^*$ states. While the integrated Raman intensity in $z$-direction is zero in position **A** due to symmetry, the signal in **B** originates mainly from the local MLCT state $D_1$, Figure 3B. In the periphery of **SnPc**, *e.g.* in **C** and **D**, mainly $\pi\pi^*$ states contribute to the resonance enhancement. Charge-transfer states are found at these tip positions (**C** and **D**) typically above 2.5 eV (Table S2). At $E_L$ = 1.959 eV, the highest integrated signals are localized between the four isoindole moieties (Figure 4B) with integrated intensities of nearly $2,500\,arb.u.$, which correlates to an enhancement the Raman signal by five orders of magnitude with respect to the non-resonant reference layer at $z$ = 20 Å. Increasing the excitation energy of the laser to 2.331 eV (532 nm) alters the nature of the contributing excited-states fundamentally. At this energy – coinciding with the surface plasmon resonance of silver nanoparticles with a diameter of approximately 50 nm[98,99] – $\pi\pi^*$ states are superimposed by bright charge-transfer states of MLCT and LMCT character. In agreement with the transition dipole maps (Table S2) featuring the highest $(\mu_z)_{ge}^2$-values in position **D**, the integrated $z$-polarized RR maps at $E_L$ = 2.331 eV exhibit the strongest signals in position **D**, while integrated signal intensities are enhanced by a factor of $6 \cdot 10^5$ (up to $\approx 17,000\, arb.u.$) – 17,000 times stronger than under non-resonant conditions within the same tip position (vdW-layer) and only due to the resonant chemical contribution. Further blue-shift of the excitation wavelengths to $\lambda_L$ = 442 nm ($E_L$ = 2.807 eV) does not affect the signal distribution along the map significantly since the same MLCT and LMCT states are in resonance as at 532 nm (2.331 eV), see Figure 4C. Thus, the most intense signals are predicted in position **D**. However, the integrated intensities are further enhanced by almost sever orders of magnitude (to nearly $226,000\, arb.u.$) as the bright charge-transfer states are in full resonance at 442-nm excitation. Consequently, the Raman intensity is amplified with respect to the non-resonant vdW-layer by more than five orders of magnitude.

Comparison of the integrated intensity maps under non-resonant and resonant conditions clearly shows that the signal at certain grid points is amplified by a factor of $10^5$ to $10^7$ upon resonance enhancement by $\pi\pi^*$ but most prominently by charge-transfer states (MLCT and LMCT). In addition, the signal contrast is increased tremendously. Hence, the integrated intensity maps point at an even increased lateral resolution under resonant conditions.

This high lateral resolution is furthermore evident from the *z*-polarized RR spectra obtained at each grid point. The normalized Raman spectra are exemplarily shown for the tip positions **A**, **B**, **C** and **D** for the three excitation wavelengths ($\lambda_L$ = 633, 532 and 442 nm), see Figure 4D-F. As stated previously, no signal is obtained at position **A** where the Raman intensity of each mode is zero by means of symmetry at all wavelengths. At 633 nm, the spectrum in **B** differs significantly form the spectra in **C** and **D**. This is reasoned by the fact that at this excitation wavelength, the $D_1$ MLCT state as well as the low-lying $\pi\pi^*$ states are in resonance in **B**, thus, this spectrum indicates an overlap of MLCT and $\pi\pi^*$ signatures localized on the central $SnN_4$ fragment. In contrast, **C** and **D** are obtained (mainly) in resonance with the low-lying pair of $\pi\pi^*$ states. The Raman spectrum in position **B** changes substantially upon variation of the laser wavelengths from 633 to 532 nm and finally to 442 nm. At 442-nm excitation, the Raman spectrum (in **B**) originates from the resonance with the weakly absorbing high-lying pair of $\pi\pi^*$ states. In contrast, less pronounced alterations of the RR intensity pattern are predicted in **C** and **D** with respect to the variation of the excitation wavelength, as the contribution of $\pi\pi^*$ states to the RR signature decreases from 633 to 442 nm. Further details regarding the vibrational modes contributing to the signal at the three excitation wavelengths are collected in Figure S2, S3 and S4 of the supporting information.

### 3.3 Tip-induced reorientation

While sections 3.1 and 3.2 address the plasmon-enhanced Raman response of the surface-immobilized **SnPc** under non-resonant and resonant conditions, section 3.3 aims to elucidate the impact of the **SnPc** orientation on the surface. Besides, the highly symmetric orientation of **SnPc** on the (fictive) surface with the central Sn atom facing downward (or upward), different orientations with the conical phthalocyanine facing towards the surface, *i.e.* by either one or two isoindole moieties, are conceivable. Both configurations lead to a symmetry reduction of the immobilized **SnPc** to $C_s$. Such reorientation is likely to be induced by close-range repulsive interactions between the tip and the sample. In the following, we focus our investigation on the edge-configuration with two isoindole fragments facing towards the supporting surface. Therefore, the **SnPc-Ag** hybrid system was rotated by 18.1° along the angle ß, see Figure 1A. Due to the reduced symmetry (from $C_{4v}$ to $C_s$), half of the sample needs to be mapped. However, at close tip-sample distances, *e.g.* at the vdW-layer, all quantum mechanically simulated properties – such as (transition) dipole moments and (transition) polarizabilities – can be obtained along the three-dimensional grid by applying the rotational matrix introduced in eq. 7 without the necessity to perform additional DFT and TDDFT calculations for the reoriented **SnPc-Ag**.

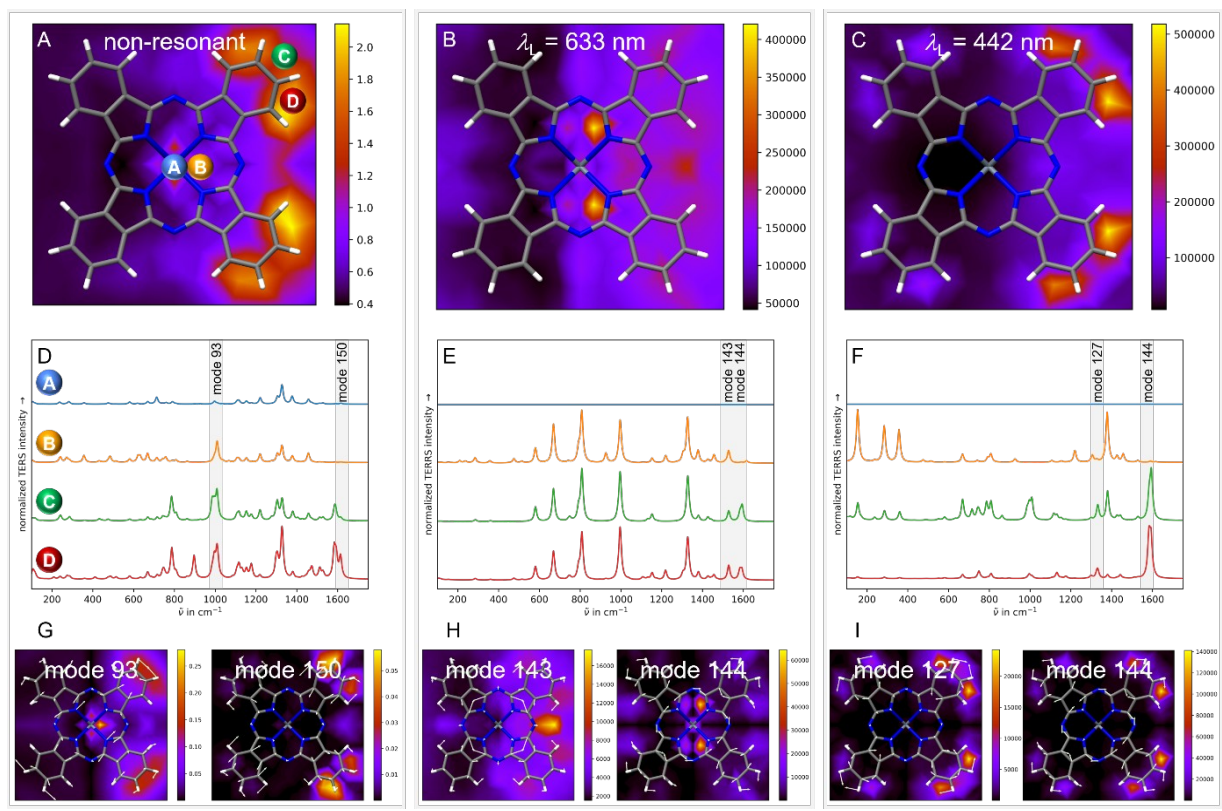

**Figure 5:** A: Integrated intensity maps of the *z*-polarized Raman signal under non-resonant and B and C: under resonant conditions ($E_L$ = 1.959 and 2.807 eV; $\lambda_L$ = 633 and 442 nm) of the hybrid system at the van-der-Waals (vdW) layer; four selected tip positions (**A**, **B**, **C** and **D**), are indicated. D-F: *z*-polarized (resonance) Raman signals at the positions **A**, **B**, **C** and **D**. G-I: Intensity maps of prominent vibrational modes contributing to the Raman spectra are illustrated in D-F.

Realignment of **SnPc** introduces pronounced variations in the integrated *z*-polarized non-resonant and resonant (or rather *z'*-polarized) intensity maps at the vdW-layer. Under non-resonant conditions, the integrated intensity map depicted in Figure 5A features, in agreement with the $C_{4v}$-orientation (Figure 2), intense signals in the region of the CH-bonds, while exclusively the signals originating from upward facing fragment are enhanced. This finding was anticipated, as the CH-centered signals mainly stem from in-plain and out-of-plain CH-bending modes. By rotation along the y-axis ($\llcorner x$ = 18.1°), the contribution of in-plain CH-bending modes is modified: Modes of the upward facing CH-groups are enhanced, while the respective modes of the CH-groups parallel to the surface are diminished, Figure 5D and G. Furthermore, and in contrast to the $C_{4v}$ orientation, the Raman signal in position **B** is less prominent for the tilted structure, while the maximum signal intensity is increased form 1 (position **B**) upon rotation along $\llcorner x$ by a factor of two (position **D**).

Resonant excitation at $E_L$ = 1.959, 2.331 and 2.807 eV ($\lambda_L$ = 633, 532 and 442 nm) leads likewise to alterations of the z-polarized Raman signals, and consequently of the integrated intensity maps, upon reorientation. Here, the fundamental difference emerges from the ratio of contributing $\pi\pi^*$ and charge-transfer (MLCT and LMCT) states. In case of the $C_{4v}$ orientation, the impact of $\pi\pi^*$ states is marginal, as these states are polarized in xy-direction with partial z-polarization induced by the silver tip at the vdW-layer. However, rotation along y-axis increases the z-component of these states by a factor of ten. This is most evident from the UV/vis absorption spectra for the surface immobilized **SnPc**, shown in Figure 3B and C. In case of the $C_{4v}$ orientation, the absorption in z-direction near the 633-nm (1.959 eV) excitation emerges from MLCT and $\pi\pi^*$ contributions, while it is governed between 500 and 400 nm mainly by charge-transfer states. In contrast, $\pi\pi^*$ ($D_1$, and $D_2$ in Figure 3C) states dominate around 650 nm for the reoriented structure, while both charge-transfer ($D_3$, $D_4$ and $D_7$ in Figure 3B) and $\pi\pi^*$ ($D_6$ and $D_7$ in Figure 3C) states contribute to the blue-sided absorption features.

Consequently, the z-polarized RR spectra upon 633-nm excitation, shown for the positions **A**-**D** in Figure 5E, feature with respect to Figure 4C enhanced contributions of the aromatic backbone of **SnPc**, *e.g.* CC-stretch and ring-breathing modes gaining z-polarization due to the reorientation. This variation is also prominent in Figure 5A, where the $\pi$-backbone is clearly visible. However, also MLCT-contributions, associated to the bright signals in-between the isoindole moieties are evident (position **D**). In agreement with the non-resonant vdW-layer, the signal intensity is generally enhanced – at $\lambda_L$ = 633 nm up to $420{,}000\,arb.u.$, which corresponds to an enhancement factor of $1.7 \cdot 10^7$.

In a similar fashion, the electronic nature at an excitation wavelength of 442 nm is shown by the integrated *z*-polarized intensity map and by the underlying Raman spectra, see Figure 5C and F, respectively. The intensity map features the most intense signal in position **D**, reflecting the contribution of the *z*-polarized MLCT and LMCT states as discussed in section 3.2. This finding is also reflected by comparison of the RR spectra shown in Figure 4F and Figure 5F. However, the integrated signal intensity in position **D** is increased from roughly $226{,}000\,arb.u.$ to $\approx 518{,}000\,arb.u.$ (enhancement: $9 \cdot 10^6 - 2 \cdot 10^7$) for the tilted orientation.

Excitation at 532 nm (not shown) can be considered as an intermediate of the previously discussed maps at 633 and 442 nm excitation, while the overall intensity is decreased due to the reduced contribution of the bright $\pi\pi^*$ states at this excitation wavelength.

**Conclusions**

In this contribution, we present for the first time a fully quantum chemical approach to disentangle the so-called chemical contributions – including ground-state interactions, charge-transfer phenomena and resonances with excited-states of the plasmonic hybrid system – in plasmon-enhanced Raman spectroscopy. Thereby, we were able to rationalize the lately postulated experimental evidence for sub-nanometer resolution of TERS under non-resonant[49] and resonant conditions[48] by means of our computational model. To this aim, the tin(II) phthalocyanine (**SnPc**) molecule was three-dimensionally "scanned" by a single silver atom, mimicking the local (chemical) interaction among the sample and the plasmonic nanoparticle.

The *z*-polarized Raman response was obtained at each grid point by DFT and TDDFT simulations.

Based on merely the chemical contribution, pronounced variations of the Raman intensity were obtained with respect to the tip position, the wavelengths of the incidental electric field as well as the orientation of the sample. Under non-resonant conditions, mainly ring breathing and out-of-plane CH bending modes were found to contribute to the *z*-polarized Raman signal. At short tip-sample distances – the region of chemical interaction between the silver tip and **SnPc** – the integrated signal Raman is amplified by a factor of up to 25 (with respect to $z \to \infty$), while certain CH bending modes are even enhanced by more than ten orders of magnitude.

Additionally, we have performed calculations when the incident field is in resonance with an electronic state of the hybrid system. Resonance with locally excited-states of **SnPc**'s $\pi$-system as well as with charge-transfer states between the tip and the sample were described based on the transform theory. It was shown that the contributions of charge-transfer and $\pi\pi^*$ states to the resonance enhancement are highly dependent on the lateral position of the silver tip, the orientation of the molecule and on the excitation wavelength. While the excitation energy of bright *z*-polarized charge-transfer states depends on the tip's position, the excitation energy of $\pi\pi^*$ states are invariant. However, an excitation of the latter **SnPc**-centered $\pi\pi^*$ states is only feasible at small tip-sample distances due to the overall small *z*-component of the respective transition dipole moments.

In addition, the impact of a tip-induced reorientation of **SnPc** on the supporting surface, *i.e.* rotation along the angle $\alpha$, on the spectroscopic response was computationally assessed. The quantum chemical calculations reveal at non-resonant excitation a moderate signal enhancement due to an increased contribution of in-plain CH stretching modes of the upward facing **SnPc** moiety. More prominent consequences regarding the local Raman signal, and thus with respect to the lateral resolution, were elucidated under resonant conditions, as the optical accessibility of the $\pi\pi^*$ states is tremendously increased. Thereby, the (integrated) signal intensity is enhanced by up to a factor of ≈10 upon reorientation. Additionally, the spectral signature of the RR signal is varied as the ratio of the contributing $\pi\pi^*$ and charge-transfer states is altered by the tip-induced reorientation. Thus, a very strong resonance enhancement of almost six orders of magnitude of the Raman signal with minute, sub-molecular modifications can be expected. Additionally, these resonance Raman spectra are not only highly sensitive to the relative position of the scanning silver tip with respect to the molecule, but also to which electronic state the incident field is resonant: charge-transfer *vs.* $\pi\pi^*$ states. As a consequence, strongly varying Raman spectra with large enhancement factors of particular vibrational modes are obtained when changing the excitation wavelength. This allows to possibly address different moieties of the molecule via directly measuring its vibrational modes.

Additionally, our independent calculations on the electromagnetic effects employing finite element methods show that for the non-resonant case, an additional enhancement of the overall Raman signal by more than one order of magnitude can be expected, resulting, together with the enhancement by the (non-resonant) chemical effect of almost three orders of magnitude, in an overall enhancement of about roughly four orders of magnitude. For the resonant cases, these numbers may vary substantially.

Future studies will focus on the incorporation of the electromagnetic effect under non-resonant and resonant conditions, the impact of further components of the polarizability tensor ($\alpha_{xx}$ and $\alpha_{yy}$) in the presence of inhomogeneous electric fields and field gradients, as well as the chemical/plasmonic effects of the supporting substrate.

## Acknowledgement

K.F., M.A., M.R. and S.G. gratefully acknowledge funding from the European Research Council (ERC) under the European's Horizon 2020 research and innovation programme – QUEM-CHEM (grant number 772676), "Time- and space-resolved ultrafast dynamics in molecular-plasmonic hybrid systems". This work was further funded by the Deutsche Forschungsgemeinschaft (DFG, German Research Foundation) – project number C2 – SFB 1375. All calculations were performed at the Universitätsrechenzentrum of the Friedrich Schiller University Jena.